\documentclass[twocolumn, pra, nofootinbib]{revtex4-1}
\usepackage[utf8]{inputenc}
\usepackage{amsmath,amsfonts,amssymb}
\usepackage{graphicx}
\usepackage{booktabs}

\begin{document}
\title{Effective Hamiltonian for doubly excited Helium states\\
 based on four dimensional Harmonic Oscillator}
\author{\surname{Torsten Victor} Zache}
\affiliation{School of  Basic Sciences, Indian Insitute of Technology, Mandi 175001}
\author{\surname{Aniruddha} Chakraborty}
\affiliation{School of  Basic Sciences, Indian Insitute of Technology, Mandi 175001}
\date{\today}

\begin{abstract}
Effective Hamiltonians for doubly excited Heliums states based on approximate $O(4)$ symmetry are revised. New quantum numbers for a 4D Harmonic Oscillator are assigned to Helium states with both electrons in the  $ n=2$ shell. An effective Hamiltonian operator is constructed and optimzed by means of non-linear least-square fits to the energy levels of the Helium isoelectronic series. The results are interpreted in terms of electron correlation and a possible unified description of two-electron atom intrashell states.
\end{abstract}
\maketitle

\section{Introduction}
The doubly excited states of Helium are still of much interest in recent research, since the effect of electron correlation is not fully understood yet. Around 1980 Herrick and Kellman \cite{kellman1980} developed a novel supermultiplet scheme to provide a better understanding of the intrashell spectrum. This model is based in approximate $O(4)$ symmetry and can be interpreted in terms of collective rotational and vibrational modes of the atom. From this "molecular" structure Kellman derived \cite{kellman1999} an effective Hamiltonian operator for the doubly excited Helium states.

In this paper, we will first follow Kellman's approach. Guided by the connection between angular momentum and two uncoupled harmonic oscillators, originally proved by Schwinger \cite{schwinger}, quantum numbers are assigned to Helium states with both electrons in the $n=2$ shell in section 3. Focusing on accuracy of the energy levels, an effective Hamiltonian operator is then constructed and its parameters are determined by performing non-linear least-square fits in section 4. This operator is also applied to similar two-electron atoms, giving some insight into the contribution of electron correlation. Section 5 summarizes the results and interprets the form of the Hamiltonian.

For further information the interested reader is referred to nice summaries of the history of collective models for two-electron atoms, given by Berry \cite{berry} and Kellman \cite{kellman1997}, and to references therein.

\section{$O(4)$ effective Hamiltonian}
\begin{figure*}
\caption{\label{fig:fits} Originial data (a) and fits of the He spectrum. 
(b) one-parameter fit with $\alpha = \beta$ and $\gamma = 0$ ;
(c) two-parameter fit with variable $\alpha , \beta$;
(d) three-parameter with variable $\alpha , \beta , \gamma$}
\includegraphics[scale=0.53]{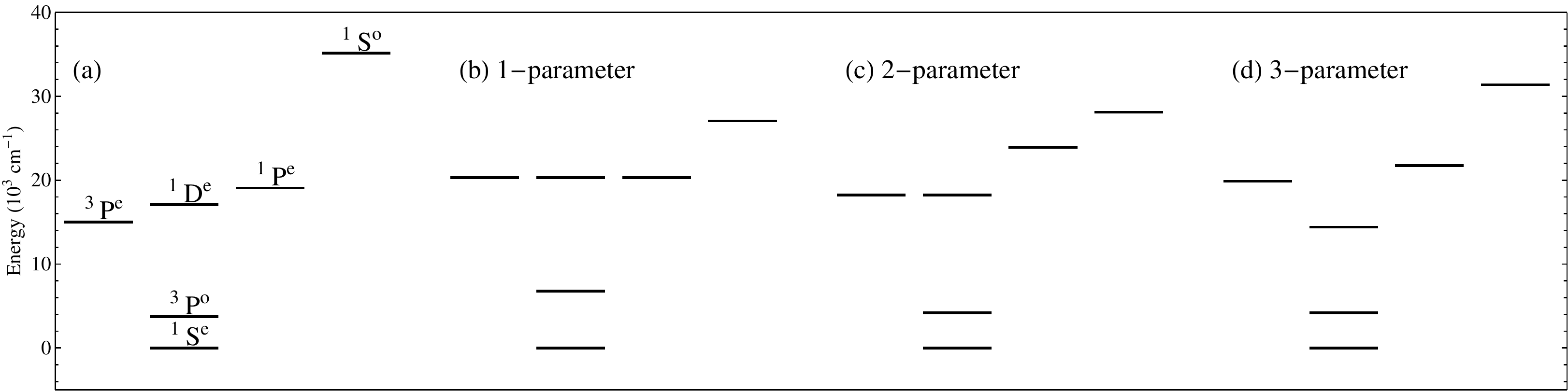}
\end{figure*}

In \cite{kellman1994} Kellman examined the origin of supermultiplets in $U(4)$ group embedding. With the resulting equations he constructed \cite{kellman1999} the following effective Hamiltonian operator (atomic units are used throughout this paper)
\begin{equation}
H_{eff} = \alpha \left( {\vec{a}_1}^2 + {\vec{a}_2}^2 \right) + \beta \vec{a}_1 \cdot \vec{a}_2 + \gamma L(L+1)
\label{eq:HO4}
\end{equation}
Here  $\vec{a}_i$ denote the Runge-Lenz operators of the single electrons, $L$ is the total angular momentum and $\alpha , \beta , \gamma$ are the parameters to be determined. The resulting equation from approximate $O(4)$ theory is \eqref{eq:HO4} with  $\alpha = \beta$. The first term in \eqref{eq:HO4} can be interpreted as a contribution by the single particles and the second as a correlation contribution. Therefore by varying $\alpha$ and $\beta$, one should be able to improve the supermultiplet scheme and achieve higher accuracy in the energy levels.
The data used for the fits is taken from \cite{lipsky} and is shown in TAB. \ref{tab:data}.
\begin{table}[b]
\caption{\label{tab:data}Energy levels ($10^3 cm^{-1}$) \cite{lipsky} for the $n=2$ states of the Helium isoelectronic series, their supermultiplet classification [P, T] and other quantum numbers}
\begin{ruledtabular}
\begin{tabular}{cccccccccc}
state & 	$e^-$ conf. (He) 	&  He 		& Li$^+$	& Be$^{2+}$	& B$^{3+}$		& P 	& T 	& L 	& A \\ 
\hline
$^{1}S^e$ 	& ${2s}^2$ 		& 0 		& 0		& 0			& 0			& 2	& 0	& 0	& 6 \\
$^{3}P^o$ 	& $2s2p$ 		& 3.719 	& 6.027	& 8.271		& 10.484		& 2 	& 0	& 1	& 4 \\
$^{3}P^e$ 	& ${2p}^2$ 		& 15.004 	& 24.455	& 33.716		& 42.886		& 1	& 1	& 1	& 2 \\
$^{1}D^e$ 	& ${2p}^2$ 		& 17.067 	& 31.024	&45.048		& 58.987		& 2	& 0	& 2	& 2 \\
$^{1}P^o$ 	& $2s2p$ 		& 19.069 	& 33.972	&49.142		& 64.329		& 1	& 1	& 1	& 4 \\
$^{1}S^e$ 	& ${2p}^2$ 		& 35.218 	& 63.287	&91.020		& 118.399		& 0	& 0	& 0	& 2 
\end{tabular}
\end{ruledtabular}
\end{table}
The first fit was done for $\alpha = \beta$ and $\gamma = 0$. Then \eqref{eq:HO4} reduces to
\begin{equation}
H_{eff}^{(1)} = \alpha \left[ P(P+2)+T^2-L(L+1) \right]
\end{equation}
where we have used (see \cite{kellman1999})
\begin{equation}
P(P+2)+T^2 = L(L+1) + \left(\vec{a}_1 - \vec{a}_2 \right)^2
\label{eq:casimir}
\end{equation}
In the approximation of independent Runge-Lenz vectors, one can use the following well-known expression from the Hydrogen atom
\begin{equation}
(\vec{a}_i)^2 + l_i(l_i +1) = {n_i}^2 - 1
\label{eq:LRL}
\end{equation}
Here $l_i$ stands for the angular momentum of a single electron and $n_i$ is the principal quantum number. In the case of $\alpha \neq \beta$ and $\gamma = 0$ and using \eqref{eq:casimir} and \eqref{eq:LRL}, \eqref{eq:HO4} becomes
\begin{equation}
H_{eff}^{(2)} = \alpha A + \beta \left[P(P+2)+T^2-L(L+1)-A \right]
\label{eq:HO4b}
\end{equation}
${A \equiv {\vec{a}_1}^2 + {\vec{a}_2}^2}$ is also shown in TAB. \ref{tab:data}.
The third fit was performed for the 3 independent parameters $\alpha ,\beta ,\gamma$.
\begin{equation}
H_{eff}^{(3)} = \alpha A + \beta \left[P(P+2)+T^2-L(L+1)-A \right] +\gamma L(L+1)
\label{eq:HO4c}
\end{equation}
We performed several non-linear least-square fits, in which the offset was determined, so that the lowest energy level was hold fixed. Consequently each fit consists of 5 independent energy levels. The results are shown in TAB. \ref{tab:results} and pictured in FIG. \ref{fig:fits}.
\begin{table}[h]
\caption{\label{tab:results}Results for the effective Hamiltonian fits ($10^3 cm^{-1}$)}
\begin{ruledtabular}
\begin{tabular}{cccc}
 			& $H_{eff}^{(1)}$ 	& $H_{eff}^{(2)}$ 	& $H_{eff}^{(3)}$ \\
\hline
$\alpha $ 		& 3.383 		& 2.090 		& 3.456 \\
$\beta $ 		& 3.383 		& 4.934		& 4.388\\
$\gamma $  		& 0			& 0			& 1.366\\
$\alpha / \beta$ 	& 1			& 0.424		& 0.788 \\
rms			& 3.480		& 2.790		& 2.403\\
\end{tabular}
\end{ruledtabular}
\end{table}

One can clearly see that the first fit reproduces the supermultiplet structure. However there is no splitting between the $^{3}P^o$ and $^{1}P^o$ states, but it can be achieved by allowing variation of $\beta$. It should be noted that the signs of $\alpha$ and $\beta$ mean, that single-particle and correlation terms enter with opposite signs in $H_{eff}$ as it is expected from theory. We will not go into further detail here, because the results obtained are basically the same as given by Kellman \cite{kellman1999}. Especially for the first fit (see FIG. \ref{fig:fits} (b)) we obtained the same numerical results, whereas the values of $\alpha , \beta , \gamma$ vary in cases (c) and (d) due to a different fitting procedure. Still the qualtiative behaviour of the Helium spectrum observed here is the same.

\section{Assignment of new quantum numbers}
\begin{figure}[b]
\caption{\label{fig:uncor}(a) Energy levels in He with single-particle contribution only; (b) the corresponding electron configuration}
\includegraphics[scale=0.5]{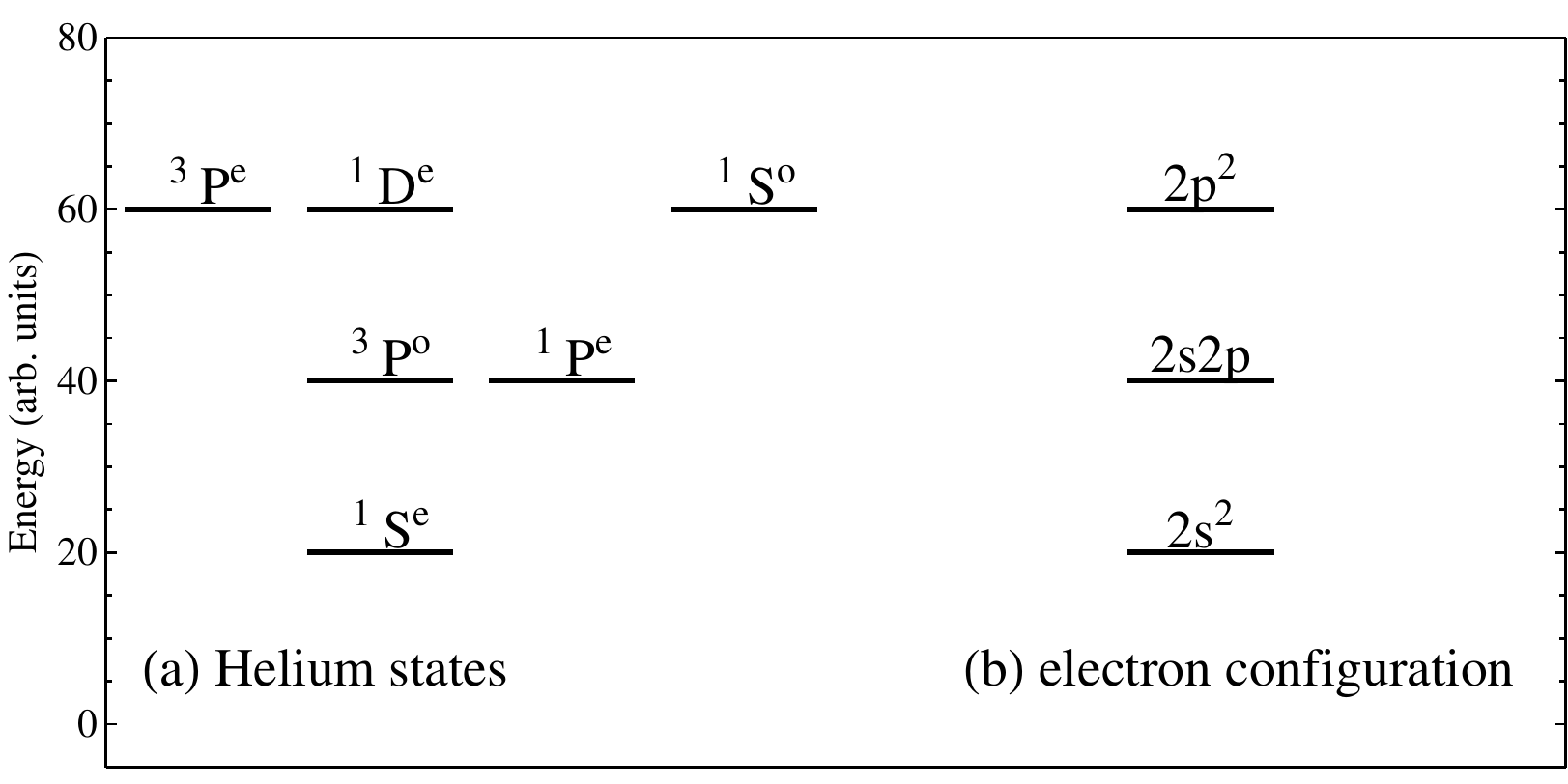}
\end{figure}

The effective Hamiltonian shown above was derived from the approximate supermultiplet structure and provides deep insight into the behaviour of the doubly excited Helium states. But even allowing variation of several parameters doesn't fit the energy levels  perfectly as can clearly be seen from the rms values in TAB. \ref{tab:results}. In the following, an empirical assignment of new quantum numbers is developed in order to improve the accuracy of the energy levels. Since this paper only deals with states where both electrons are in the $n=2$ shell, only four quantum numbers are needed to describe these. Schwinger proved \cite{schwinger} that it is possible to express any angular momentum with quantum numbers $l,m$ through two uncoupled harmonic oscillators ("HO") with quantum numbers $n^{(1)}, n^{(2)}$, so that
\begin{eqnarray}
l &= \frac{1}{2} \left(n^{(1)}+ n^{(2)} \right) \label{eq:l}\\
m &= \frac{1}{2} \left(n^{(1)}- n^{(2)} \right) \label{eq:m}
\end{eqnarray}
Therefore we tried two construct a pair of HO for each electron starting from the uncorrelated case in which there are only three different energy levels, depending on the electron configuration (see FIG. \ref{fig:uncor})
Upon postulating, that the Hamiltonian operator to be developed in the following section should consist of a HO part, that yields the uncorrelated structure in first order, we arrive at the following equations for the quantum numbers $n_s^{(1)}, n_s^{(2)}$ and $n_p^{(1)}, n_p^{(2)}$ for electrons in $s$- and $p$- states respectively.
\begin{widetext}
\begin{eqnarray}
\text{2s}^2 &: 2a + \text{const}	=&\left( n_s^{(1)}+\frac{1}{2} \right)+\left( n_s^{(2)}+ \frac{1}{2} \right)+\left( n_p^{(1)}+\frac{1}{2} \right)+\left( n_p^{(2)}+ \frac{1}{2} \right) \nonumber\\
\text{2s2p} &: 3a + \text{const}	=&\left( n_s^{(1)}+\frac{1}{2} \right)+\left( n_s^{(2)}+ \frac{1}{2} \right)+\left( n_p^{(1)}+\frac{3}{2} \right)+\left( n_p^{(2)}+ \frac{3}{2} \right)\nonumber\\
\text{2p}^2 &: 4a + \text{const}	=&\left( n_p^{(1)}+\frac{3}{2} \right)+\left( n_p^{(2)}+ \frac{3}{2} \right)+\left( n_p^{(1)}+\frac{3}{2} \right)+\left( n_p^{(2)}+ \frac{3}{2} \right) \nonumber
\end{eqnarray}
\end{widetext}
Here $a$ scales the energy gap between the levels of FIG. \ref{fig:uncor} and is related to the angular momentum, but the latter interpretation shouldn't be taken too seriously. It is to be thought of as the starting point of this model, but will lead to some problems (see section 5). Chosing the arbitrary constant, so that $a$ and $n_{s/p}^{(i)}$ become integer values (with minimal HO quantum numbers), we arrive at
\begin{eqnarray}
 n_s^{(1)} +  n_s^{(2)} 	&= 0 \nonumber\\
 n_p^{(1)} +  n_p^{(2)}	& = a \nonumber
\end{eqnarray}
The obvious solution is $ n_s^{(1)} = n_s^{(2)} = 0$, whereas $ n_p^{(i)}$ depends on $a$:
\[  [n_p^{(1)}, n_p^{(2)}] \in \{ [0,a],[1,a-1], \dots , [a-1,a],[a,0]\}\]
$a=0$ doesn't yield different energy levels. For $a=1$, the following states ($[n_a^{(1)}, n_a^{(2)},n_b^{(1)}, n_b^{(2)}]$ for electrons $a,b$) are derived 
\begin{eqnarray}
\text{2s}^2 :& [0,0,0,0]\nonumber \\
\text{2s2p} :& [0,0,0,1] , [0,0,1,0] \nonumber \\
\text{2p}^2 :& [0,1,0,1],[1,0,0,1],[1,0,1,0],[0,1,1,0] \nonumber
\end{eqnarray}
This is a total number of 6 different states, since [0,1,1,0] and [1,0,0,1] represent the same state due to indistinguishability of the electrons. However if the two HO quantum numbers of each electron should be treated equal, which is a reasonable assumption, it is impossible to construct a Hamiltonian operator where the $2s2p$ states $^{3}P^o$ and $^{1}P^o$ behave differently. Therefore we conclude, that [0,1] and [1,0] represent the same state\footnote{for a single electron} and choose to display the state with quantum numbers of increasing magnitude. Finally in case of $a=2$, the possible states are:
\begin{eqnarray}
\text{2s}^2 :& [0,0,0,0]\nonumber \\
\text{2s2p} :& [0,0,1,1] , [0,0,0,2] \nonumber \\
\text{2p}^2 :& [0,2,0,2],[1,1,0,2],[1,1,1,1] \nonumber
\end{eqnarray}
The energy of a state is given by the sum of all four quantum numbers and therefore coincides with the structure of FIG. \ref{fig:uncor} by construction. We chose the assignment to original states given in TAB. \ref{tab:qn}, keeping in mind their original ordering in TAB. \ref{tab:data}. Comparison with TAB. \ref{tab:qn} allows a first interpretation\footnote{Actually we have \emph{chosen} the labels, so that this behaviour is observed. We could have exchanged the meaning of [1,1] and [0,2], but then of course a different Hamiltonian would have been necessary to fit the energy levels} of [1,1] as lowering and [0,2] as lifting the energy level. So far it looks like there are two different kind of excitations of electrons resulting in a different correlation energy.
\begin{table}[h]
\caption{\label{tab:qn}Assignment of the 4D HO quantum numbers to the doubly excited He states}
\begin{ruledtabular}
\begin{tabular}{rlc}
state &electron conf. 	&$[n_a^{(1)}, n_a^{(2)},n_b^{(1)}, n_b^{(2)}]$  \\ 
\hline
$^{1}S^e$ & 2s$^2$ 	& [0,0,0,0] \\
$^{3}P^o$ & 2s2p		& [0,0,1,1] \\
$^{3}P^e$ & 2p$^2$	& [1,1,1,1] \\
$^{1}D^e$ & 2p$^2$	& [1,1,0,2] \\
$^{1}P^o$ & 2s2p		& [0,0,0,2] \\
$^{1}S^e$ & 2p$^2$	& [0,2,0,2] 
\end{tabular}
\end{ruledtabular}
\end{table}

\section{Effective Hamiltonian for 4D Harmonic Oscillator}
\begin{figure}[h]
\caption{\label{fig:corr}Contribution of the electron correlation; $a=0.107$, $b=1.302$, $c=0.349$}
\includegraphics[scale=0.3]{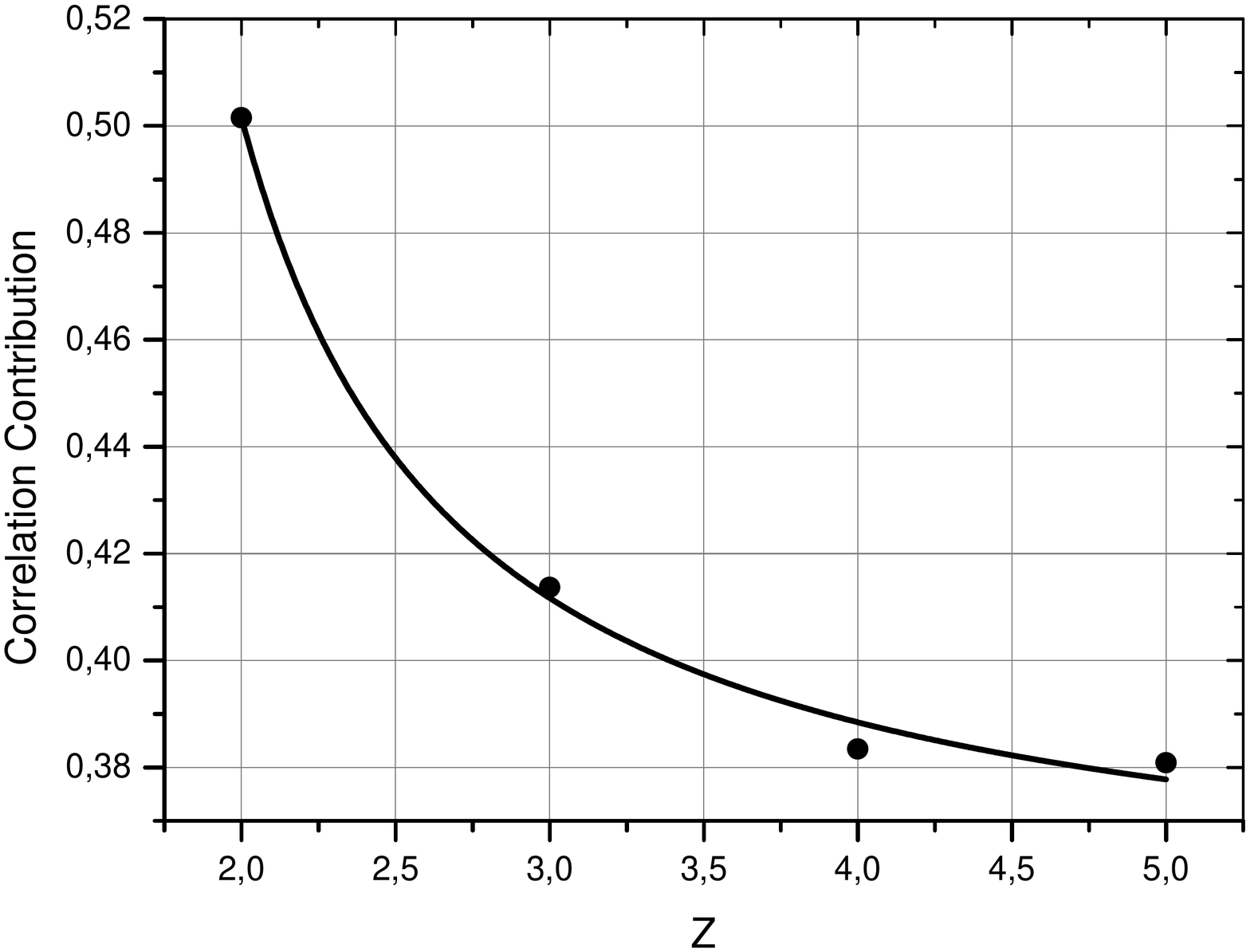}
\end{figure}

\begin{figure*}
\caption{\label{fig:4DHO}\emph{black:} original energy levels; \emph{green (light grey):} result of the fits}
\includegraphics[scale=0.7]{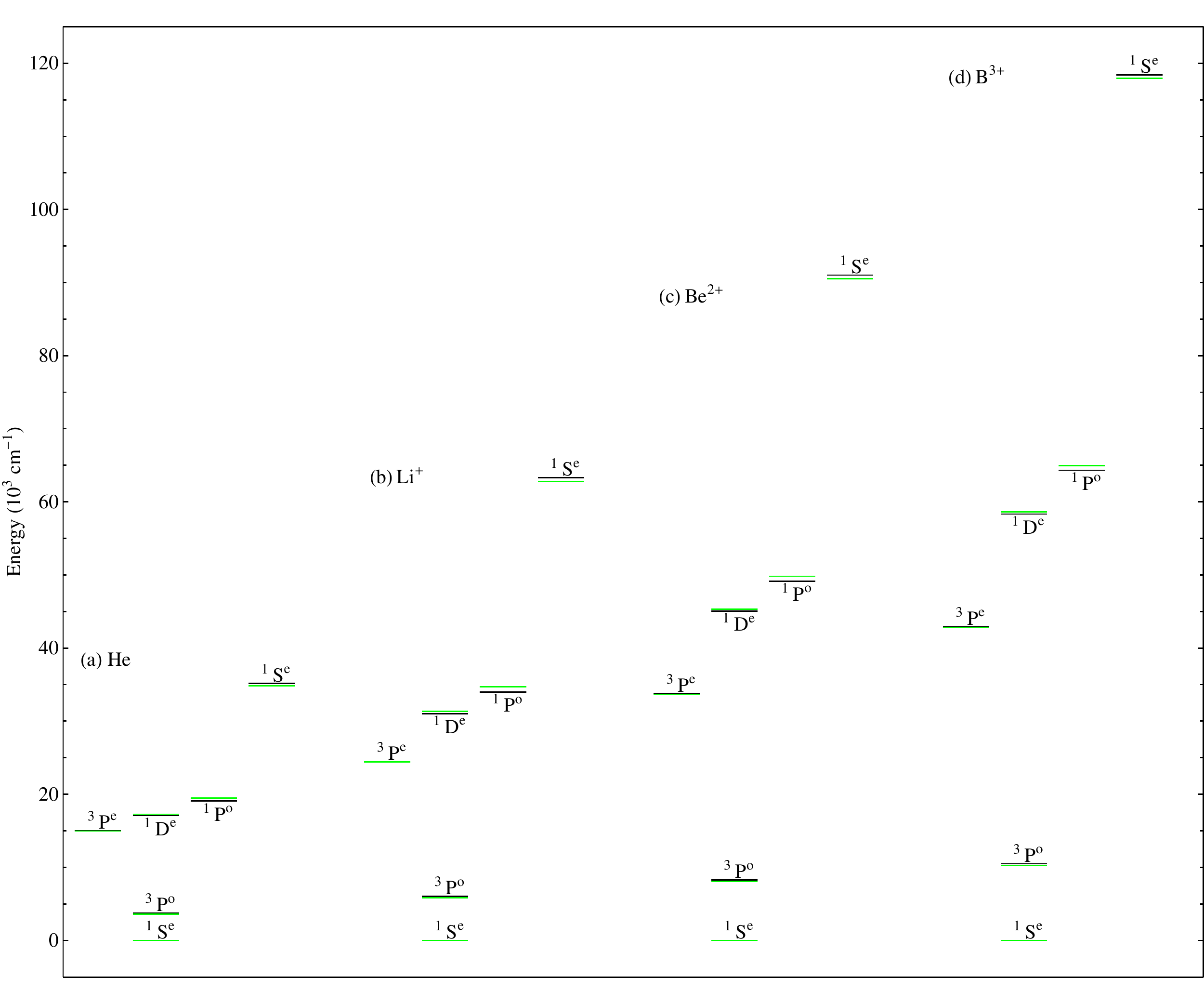}
\end{figure*}

In this section we give another effective Hamiltonian operator based on the four dimensional HO scheme developed above and optimize its parameters by means of non-least square fits. This operator consists of a HO term $H_{HO}$ and an interaction term $H_{int}$. The HO part again consists of a first and second order term, whereas the interaction term expresses two different kinds of interaction.
\begin{eqnarray}
H 	=& H_{HO} &+ H_{int} \label{eq:H}\\
	\equiv & \left( \alpha  H_0 + \beta H_1 \right) &+ \left(\gamma  H_{int}^{(1)} + \delta H_{int}^{(2)} \right) \nonumber
\end{eqnarray}
\begin{table}[h]
\caption{\label{tab:fit}Result of the fit in $10^3 cm^{-1}$}
\begin{ruledtabular}
\begin{tabular}{ccccc}
 		&He 		& Li$^+$	& Be$^{2+}$	& B$^{3+}$  \\
\hline
$\alpha$	& -8.845 	& -17.549	& -26.1523		& -34.433 \\
$\beta$	& 6.442	& 12.020	& 17.551		& 22.995 \\
$\gamma$	& 3.581 	& 5.720	& 7.837		& 10.217 \\
$\delta$	&-4.086 	& -6.512	& -8.921		& -11.657\\
rms		& 0.060	& 0.162	& 0.138		& 0.122
\end{tabular}
\end{ruledtabular}
\end{table}
\begin{widetext}
\begin{eqnarray}
H_0=& n_a^{(1)}+ n_a^{(2)}+n_b^{(1)}+ n_b^{(2)}+2 \label{eq:H0}\\
H_1=&\left( n_a^{(1)}+\frac{1}{2} \right)^2+\left( n_a^{(2)}+ \frac{1}{2} \right)^2+\left( n_b^{(1)}+\frac{1}{2} \right)^2+\left( n_b^{(2)}+ \frac{1}{2} \right)^2 \label{eq:H1}\\
H_{int}^{(1)}=& \left( n_a^{(1)}+\frac{1}{2} \right)^2 \left( n_a^{(2)}+ \frac{1}{2} \right)^2 \left( n_b^{(1)}+\frac{1}{2} \right)^2 \left( n_b^{(2)}+ \frac{1}{2} \right)^2 \label{eq:Hint1}\\
H_{int}^{(2)} =&\left( n_a^{(1)}+\frac{1}{2} \right)^2 \left( n_a^{(2)}+ \frac{1}{2} \right)^2 \left( n_b^{(1)}+\frac{1}{2} \right) \left( n_b^{(2)}+ \frac{1}{2} \right) +
			 \left( n_a^{(1)}+\frac{1}{2} \right) \left( n_a^{(2)}+ \frac{1}{2} \right) \left( n_b^{(1)}+\frac{1}{2} \right)^2 \left( n_b^{(2)}+ \frac{1}{2} \right)^2 \label{eq:Hint2}
\end{eqnarray}
\end{widetext}
This operator is invariant under exchange of the electrons and also treats the two HO for each electron equally. The second order HO is taken in addition to the first order term, because the 4D HO model is expected to hold only approximately. By varying the four parameters we look for an optimal representation for the Helium energy levels. From theoretical background $\alpha$ is expected to take a greater value than $\beta$, otherwise the HO approximation would fail. We also expect a different sign for $\alpha$ and $\beta$ as is usually the case for second order corrections. The two interaction terms have been chosen empirically. Their parameters are expected to take a smaller value than those of the HO terms, but they won't be negligibly small, because the electron interaction plays an important role in the doubly excited Helium states. As before, the lowest energy is hold fixed, which leaves only 5 independent energy levels for 4 parameters. The fit is also applied to other atoms in the Helium isoelectronic series. The result of the fit is given in TAB. \ref{tab:fit}. It is impressive, that the rms value of the fit (especially for He) is very small, consequently the energy levels as seen in FIG. \ref{fig:4DHO} are very accurate. We also observe the postulated behaviour of the parameters. Additionally, $\gamma$ and $\delta$ have different signs, which shows that the two different forms of interaction contribute to lifting and lowering the energy levels. 
Looking only at single energy levels, we observe a linear dependence on the charge of the nucleus $Z$ in FIG. \ref{fig:4DHO}. It is interesting to find, that this dependence reappears in every fitting parameter. To demonstrate the contribution of the interaction term  in \eqref{eq:H}, we have plotted
\[\frac{| \gamma | + | \delta |}{| \alpha |+| \beta |} \]
as a function of $Z$ in FIG. \ref{fig:corr}. It becomes clear, that the effect of electron correlation gets less important for atoms with higher $Z$. The dependence can be approximated very well by a function of the form
\[ f(Z) = \frac{a}{Z-b} + c\]
The parameters are given in FIG. \ref{fig:corr}. From the exact Hamiltonian, one would expect $f \sim \frac{1}{Z}$. So there could be more effects included here. For example we can't quantify to what extend correlation effects are already included in the new labeling of the states, which could possibly lead to an effective charge $Z_{eff} = Z - b$.

\section{Summary and discussion}
In this paper, we have started to extend Kellman's work on effective Hamiltonians for doubly excited states of the Helium atom. We started with his approach \cite{kellman1999} and revised his method. The results obtained are basically the same. In the next step, we were guided by a connection between angular momentum and two uncoupled harmonic oscillators to assign new quantum numbers of a four dimensional harmonic oscillator to the $n=2$ intrashell Helium states. From this HO model we constructed an effective Hamiltonian operator and optimized it by numerical fitting to the original energy levels. Although the results are very accurate, one has to be very careful interpreting this model.

First of all, the fit was performed in order to optimize four parameters, but only five energy levels were independent, which makes the fit not very reliable. Secondly, interpretation of the quantum numbers as a usual harmonic oscillator is dangerous. The connection made by Schwinger \cite{schwinger} leads to equations \eqref{eq:l} and \eqref{eq:m}. From \eqref{eq:l} one recovers the angular momentum of the single particles, therefore leading to the uncorrelated energy levels. This was the starting point for assigning the quantum numbers. However, if our assignment would be exact, \eqref{eq:m} states that each state has a well defined magnetic quantum number, which is obviously not the case! This underlines fact, that these "quantum numbers" are only approximations. The corresponding operators do not commute with the Hamiltonian, therefore it might be better to talk about new "labels".

Furthermore the empirical nature of our model becomes apparent, since it only fits the energy levels well, when the interaction terms in \eqref{eq:H} are taken into account as well. Another point to mention is the degeneracy of the energy levels, which does not coincide with the model of a four dimensional harmonic oscillator.

At this point the question arises, whether it is possible to extend the model presented here to other states of doubly excited two-electron atoms. At first the assignment of quantum numbers seems very much limited to the special problem of $n=2$ intrashell states presented here. However, the following procedure also allows a description of $n=3$ states. For $a=0$, there is only one possibility for four harmonic oscillators, namely $[0,0,0,0]$. We interpreted this as the 2s$^2$ state, but it could also stand for $n$s$^2$ states in general.\footnote{We also mention that the $n=1$ intrashell state is described trivially.} $a=2$ additionally produced two 2s2p states and three 2p$^2$ states, which could also be interpreted as general $n$s$n$p and $n$p$^2$ states. In the same way $a=4$ could be used to construct $n$s$n$d, $n$p$n$d and $n$d$^2$ states. It would be very interesting to see, how well higher intrashell states can be described this way.

We would like to give some more thoughts about the exact form of the constructed Hamiltonian operator. One might wonder, why the first interaction term \eqref{eq:Hint1} is of second order. We also considered a term of first order, but its contribution to the energy was relatively small compared to the second order term. In order to keep the number of parameters low, we left out this term. The second interaction term \eqref{eq:Hint2} mixes vibrational contributions of different order and contributes against the first interaction. This combination is essential to achieve the correct energy levels, since using only one form of interaction tends to over- or underestimate the contribution of electron correlation.

Additionally, the fact that the operator is based on purely vibrational modes (appearing as $n + \frac{1}{2}$) is very interesting, since previous research \cite{kellman1997} points in the direction of a "molecular" structure of Helium-like atoms with collective vibrational and rotational contributions. Theoretically our modell could also be applied to other two-electron atoms, but these appear to fit the "molecular" model even better than Helium \cite{kellman1985}. More intersting is the expansion of our work to three-electron atoms.This could  lead to a purely vibrational Hamiltonian operator, while a molecular model for three-electron-atoms has not been found yet. Fitting their energy levels could give some new insight into the interpretation of collective and single-particle contributions.

In conclusion, we can say, that the model developed in this paper does provide a very nice description of the doubly excited Helium states with both electron in the $n=2$ shell. However, one has to be very careful interpretating the result, since the assignment of quantum numbers and the construction of the effective Hamiltonian operator is not based on a rigorous theoretical model, but is rather empirical. The expansion of the model to other intrashell spectra or more complicated systems, like three-electron atoms, remains the subject of further research.

\begin{acknowledgements}
Some of the figures for this article have been created using the LevelScheme scientific figure
preparation system [M. A. Caprio, Comput. Phys. Commun. 171, 107 (2005),
http://scidraw.nd.edu/levelscheme].

We would like to thank the German Academic Exchange Service (DAAD) for providing scholarship for Torsten Zache during his internship at IIT Mandi.
\end{acknowledgements}

\bibliography{ref}

\end{document}